\documentclass{article}
\usepackage{dcase2023,amsmath,graphicx,url,times,booktabs, tabularx, lipsum}

\title{Leveraging Geometrical Acoustic Simulations of Spatial Room Impulse Responses for Improved Sound Event Detection and Localization}

 \name{Christopher Ick \hspace{1cm} Brian McFee\thanks{This material is based upon work supported by the National Science Foundation under NSF Award 1922658.}}
 \address{Music and Audio Research Laboratory, New York University\\
       Brooklyn, NY 11201, United States}
\begin{document}

\ninept
\maketitle

\begin{sloppy}

\begin{abstract}
As deeper and more complex models are developed for the task of sound event localization and detection (SELD), the demand for annotated spatial audio data continues to increase.
Annotating field recordings with 360$^{\circ}$ video takes many hours from trained annotators, while recording events within motion-tracked laboratories are bounded by cost and expertise.
Because of this, localization models rely on a relatively limited amount of spatial audio data in the form of spatial room impulse response (SRIR) datasets, which limits the progress of increasingly deep neural network based approaches.
In this work, we demonstrate that simulated geometrical acoustics can provide an appealing solution to this problem.
We use simulated geometrical acoustics to generate a novel SRIR dataset that can train a SELD model to provide similar performance to that of a real SRIR dataset.
Furthermore, we demonstrate using simulated data to augment existing datasets, improving on benchmarks set by state of the art SELD models.
We explore the potential and limitations of geometric acoustic simulation for localization and event detection.
We also propose further studies to verify the limitations of this method, as well as further methods to generate synthetic data for SELD tasks without the need to record more data.
\end{abstract}


\section{Introduction}
\label{sec:intro}

Sound event detection and localization (SELD) is the union of two active fields of research; sound event detection (SED), and localization, or direction-of-arrival (DoA) estimation.
Expanding the scene-description capabilities of SED with the spatiotemporal characterization of localization sees applications ranging from autonomous robot navigation \cite{bondi} and urban monitoring \cite{stevensspatial}, to speaker diarization \cite{wangspatial} and immersive experiences in virtual and augmented reality devices.

While earlier techniques for SELD have focused on traditional signal-processing or parametric models such as \cite{param1,param2,param3,params4}, recent literature is dominated by deep neural network (DNN) approaches, which have been shown superior performance in both pure localization \cite{doa1, doa2} as well as joint SELD tasks \cite{seldnet}.
A surge of interest in this field can be attributed to the introduction of SELD as a task in the DCASE2019 challenge \cite{dcase2019}.
This challenge included the release of several 4-channel audio datasets with spatial and temporal annotations for sound events.
The audio was generated by convolving sound events with spatial room impulse responses (SRIRs) recorded in 5 separate rooms at 504 unique azimuth-elevation-distance combinations.
This was further iterated upon by the SELD challenge in DCASE2020 \cite{DCASE2020} with 13 unique rooms.
Recently, DCASE2022 reintroduced the challenge with hand-annotated real-world recordings in the STARSS22 dataset \cite{starss22}, providing one of the first datasets with real-world data upon which to evaluate SELD models.
This dataset uses a combination of 360${^\circ}$ video, and motion capture to extract spatiotemporal annotations that were manually validated.
In addition to this, the DCASE2022 dataset also included a release of the SRIRs used to generate the training data for the SELD task, as well as the code for the generator itself, allowing users to generate their own annotated spatial data \cite{tausrir}. This data included SRIRs measured over a wide range of positions over 9 different rooms in Tampere University's campus. 

These datasets are unique in the density of SRIR measurements across particular paths, the variety of acoustic enclosures, and the large amount of SRIRs.
Because of their scale, visibility, and quality, these datasets have become some of the most cited SRIR datasets for DNN-based approaches to SELD, because of their ability to meet the data requirements of these highly-parametrized models.
Despite this, these datasets are still severely limited by the recording procedure of SRIRs, which require time, expertise, and a low-noise environment to produce at a high quality.
Increasing the spatial density, variety of trajectories types, and number of trajectory paths becomes multiplicatively time consuming to develop.
Furthermore, the range of rooms in which these measurements can be recorded is inherently limited by the limitations of the recording facilities, usually limited to a dozen rooms or so in the best of cases.
However, without a wide range of acoustic environments to perform these measurements, generalization to a variety of unseen acoustic environments becomes impossible.

Physical acoustic simulations provide an attractive solution to the limitations of field-recorded SRIRs.
Acoustic simulation is typically split into two categories: wave based methods, which simulate the propagation of sound waves through physical media, and geometric modeling methods, which model the transportation of acoustic energy through acoustic rays, mimicking popular methods for modeling optical rays.
Geometrical acoustics approximate the wavelength of the propagating sound to have wavelength relatively small compared to the room geometries of interest, and neglects wave effects such as diffraction or scattering.
Nevertheless because of ease of implementation and computational efficiency, geometrical acoustic modeling methods have seen wide success in several tasks, including modeling architectural acoustics \cite{smallroom} and room parameter estimation \cite{sonos}.

In this work, we propose utilizing one method of geometrical acoustics modeling, the image source method, to generate simulated SRIRs for training DNN models for SELD.
We demonstrate the effectiveness of this simulated method for SRIR generation using the framework and data provided in previous DCASE SELD challenges.
By creating an audio dataset from simulated SRIRs, we train a SELD model with similar performance to one utilizing real-world SRIRs.
By directly compares simulated SRIRs with a datasets of recorded SRIRs of the similar size, room geometries, and DoA distributions, we demostrate the downstream effects of simulation in place of recording as being relatively minimal, differing our work from prior studies \cite{synth-doa}.
Furthermore, we augment a typical SRIR dataset with simulated SRIRs, training models that outperform those trained solely on recorded SRIRs.
Finally we propose further experiments to explore the use of simulated SRIRs for training SELD models.
The code associated with this work is released in an open-source github repository \footnote{\url{https://github.com/ChrisIck/DCASE_Synth_Data}} to further work in using synthetic SRIRs for training DNN models.

\section{Acoustic Simulations}
\label{sec:sim}

\subsection{The Image Source Method}
\label{ssec:ism}

The image source method (ISM) is a technique used in architectural acoustics and room modeling to predict the sound field in enclosed spaces \cite{ISM}.
The ISM considers the primary sound source and virtual images reflected by the room's boundaries.
These virtual sources are assumed to emit sound with the same magnitude and phase as the primary source, but with a delay due to the additional path length traveled.
Typically, this starts with defining the room geometry, including the positions and shapes of the walls, ceiling, and floor.
For each reflecting surface, virtual image sources are ``mirrored'' across the boundary.
The number of virtual sources depends on the order of reflections considered.
From here, the interaction between the primary sound source and the virtual image sources is calculated by determining the path lengths, time delays, and attenuation factors associated with each source-receiver combination, as well as the material properties of each surface through which the sound path is reflected upon.
By summing the contributions of the primary source and its image sources, the sound field at various locations in the room can be predicted, providing an estimation of the sound pressure level, arrival times, and directivity patterns.

It's important to note several limitations of the ISM model.
The ISM implicitly assumes that all surfaces are perfectly reflective and flat, with idealized acoustic properties, which fails to account for acoustic effects such as scattering or diffraction.
Furthermore, the order of reflections/virtual sources scales the computational cost exponentially, meaning compute late-stage reflections in an SRIR prohibitively expensive.

Despite its limitations, the image source method is widely used due to its computational efficiency and effectiveness in predicting sound fields in enclosed spaces.
Because localization is more reliant on direct sounds/early reflections, the limitations caused by use of the the ISM for computing SRIRs can be expected to be relatively minimal.

\subsection{The TAU-SRIR Dataset}
\label{ssec:tau-srir}

To validate the use of ISM-generated SRIRs in a direct-comparison, we take the existing TAU-SRIR database \cite{tausrir} as an example database for which well established metrics for SELD have been measured.

The TAU-SRIR database contains SRIRs recorded in 9 different rooms throughout Tampere University's campus.
Each SRIR is computed by recording a maximum-length sequence (MLS) played through a loudspeaker, recorded on an Eigenmike spherical microphone array.
Each SRIR was downsampled to 24kHz and truncated at 300ms, resulting in $7200$ samples per RIR.
The data is stored in a 4-channel audio corresponding to a tetrahedral microphone array with the geometry in spherical coordinates $(\phi, \theta)$, specified in Table \ref{tab:mic_geom}. For each room, the position of the microphone array was provided.

\begin{table}
    \centering
        \begin{tabular}{c|rr}
        & Azimuth ($\phi$) & Elevation ($\theta$) \\
        \hline
         M1 & $45^\circ$ & $35^\circ$\\
         M2 & $-45^\circ$ & $-35^\circ$\\
         M3 & $135^\circ$ & $-35^\circ$\\
         M4 & $-135^\circ$ & $35^\circ$
        \end{tabular}
    \caption{Microphone Geometry for TAU-SRIR dataset. Each microphone is 4.2cm from the center, and is modeled with a hypercardioid response.}
    \label{tab:mic_geom}
\end{table}

SRIRs were measured along either circular or linear traces at fixed distance from the microphone array along the z-axis at a number of trajectory groups, separated by distance and reflection across the axis of the microphone array in the case of linear traces.
Circular trajectory groups had a specified radius of orbit, whereas linear trajectories had a specified start and end point in 3D space.
Each trajectory was repeated at a number of different heights, and each trajectory had a fixed number SRIR measurements and corresponding DoA measurements recorded as Cartesian components of a unit vector.
The number of SRIR measurements vary across different trajectories/heights, spaced in roughly $1^{\circ}$ increments.
The total number measurements can be seen in Table \ref{tab:room_info}

\begin{table}
    \centering
    \begin{tabular}{l|llll}
        Room Name    & Traj. type & $N_t$ & $N_h$ & $N_{\text{SRIRs}}$ \\
        \hline
        Bomb shelter & Circular   & 2              & 9         & 6480         \\
        Gym          & Circular   & 2              & 9         & 6480         \\
        PB132        & Circular   & 2              & 9         & 6480         \\
        PC226        & Circular   & 2              & 9         & 6480         \\
        SA203        & Linear     & 6              & 3         & 1594         \\
        SC203        & Linear     & 4              & 5         & 1592         \\
        SE203        & Linear     & 4              & 4         & 1760         \\
        TB103        & Linear     & 4              & 3         & 1184         \\
        TC352        & Circular   & 2              & 9         & 6480        
\end{tabular}
    \caption{Trajectory information for rooms contained in the TAU-SRIR dataset \cite{tausrir}. Each room contains trajectories across a number of trajectory groups ($N_t$) and a number of heights ($N_h$), for a total of $N_t \times N_h$ trajectories per room. Each trajectory is sampled in roughly $1^{\circ}$ increments.}
    \label{tab:room_info}
\end{table}

\subsection{Room Simulation}
\label{ssec:room-sim}
\begin{figure}[h]
    \centering
    \includegraphics[width=.45\textwidth]{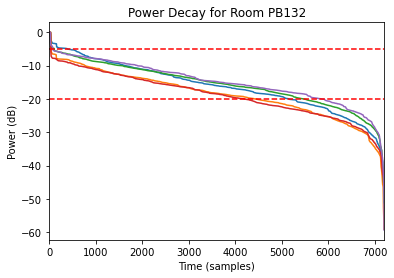}
    \caption{The log-scale energy decay from a sample of RIRs from a singular room. The region in the dashed lines is roughly linear, suggesting it corresponds to mid-late reflections, and is used for an estimate of RT$_{60}$}
    \label{fig:rir-decay}
\end{figure}

We recreate this dataset using the python package \textit{pyroomacoustics} \cite{pyroomacoustics}, a pythonic implementation of the ISM, that has demonstrated use in implementations of various algorithms for beamforming, direction finding, adaptive filtering, source separation, and single channel denoising.

To replicate the acoustic conditions of each of the rooms in the TAU-SRIR dataset, we randomly sampled the RIRs uniformly in each room until we had a sample of 5 single-channel RIRs.
Using the Schroeder method \cite{schroeder}, we estimated the RT$_{60}$ of each room by hand-selecting the early decay of the energy-decay function of the RIR samples and computing the linear fit.
This can be see in in Figure \ref{fig:rir-decay}.
Using the inverse Sabine formula, we used this to estimate the mean absorption coefficient of the rooms and the number of required reflection orders to approximate a room of a similar RT$_{60}$.
We combined these parameters with the geometry estimations from the TAU-SRIR dataset to construct virtual rooms matching those of the 9 rooms in the TAU-SRIR dataset.
To this room, we added a virtual tetrahedral microphone with the geometry specified in Table \ref{tab:mic_geom}, with each virtual microphone using a hypercardioid response pattern centered at the position specified in the TAU-SRIR dataset.

\begin{figure}[h]
    \centering
    \includegraphics[width=0.35\textwidth]{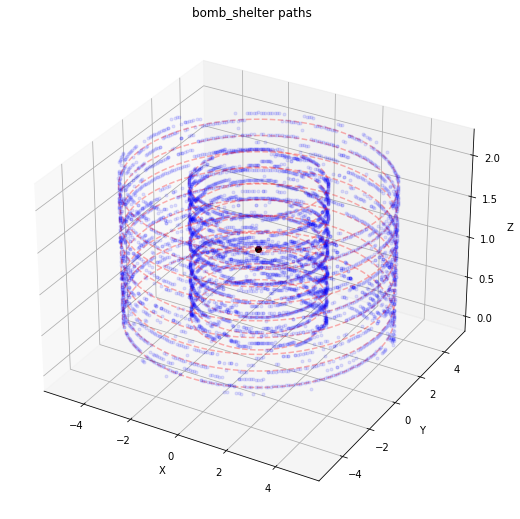}
    \caption{SRIR measurement positions reconstructed from the DoAs provided in the TAU-SRIR dataset (blue), compared to the path specified by the height, radius, and position labels provided (red)}
    \label{fig:bs-paths}
\end{figure}

To estimate the positions of the SRIRs in space corresponding to the DoA measurements in the TAU-SRIR dataset, we chose points along the DoA that most closely matched the corresponding path on the line specified by the trajectory in the dataset.
Circular datasets specified their height and radius, and were centered along the same z-line as the microphone array.
Linear trajectories specified the start point and end points of their traces.
These matched points were estimated by projecting the DoA vectors onto a cylinder that matched the radius of the trajectory, after translation by the position of the microphone array and the height of the trajectory of interest (see Figure \ref{fig:bs-paths}).
Once these points were estimated, they were placed into the simulated room in a position relative to the virtual microphone array, and the 4-channel SRIR was computed with the ISM for each point, at the sample rate of 2400kHz. 
To match the dimensions of the original TAU-SRIR dataset, we truncate the SRIRs to 300ms, providing us with an SRIR with dimensions of $7200 \times 4$ for each point.

\section{Methodology}
\label{sec:methods}
To evaluate the performance of this dataset in SELD tasks, we generated a dataset of audio events consistent with the methodology of dataset generation for training the baseline SELD model in the DCASE 2022 challenge \cite{Shimada2022}.
We generated 3 datasets, one using the original SRIR database, which we will refer to as the TAU-SRIR datset. We also generated a dataset using only the synthetic SRIRs, which we refer to as the SIM-SRIR dataset.
Finally, we generated a third dataset that equally samples both the original and simulated SRIRs, which we will refer to as the augmented SRIR dataset, or AUG-SRIR.

\subsection{Data Generation}
\label{ssec:datagen}
To generate our annotated spatialized audio, we followed the procedure used in DCASE2019-2021, by convolving various sound events with SRIRs.

The sound events were drawn from the FSD50k audioset \cite{fsd50k}, a subset containing over 20k sound events of 13 classes selected for the DCASE challenge.
These sound events were spatialized into virtual recordings, each corresponds to a singular room, allowing for up to three concurrently active sources.
The sources can be static or dynamic, with equal probability, and the dynamic sources can move at slow ($10^\circ$/sec), moderate ($20^\circ$/sec), or fast ($40^\circ$/sec) angular speeds.
Each sample lasted 60 seconds, 40 of which had at least one active event class.

For each of the 3 SRIR datasets, 1200 recordings were created in separate folds, 900 for training and 300 for validation.
The training and validation sets used 6 and 3 rooms respectively, such that none of the same rooms overlapped both folds.

\subsection{Model}
\label{ssec:model}
The model architecture is identical to the one used in the DCASE2022 Task 3 challenge baseline \cite{starss22}; a SELDnet style CRNN with multi-ACCDOA representation for co-occuring events \cite{multiaccdoa}.
The model takes $T$ frames of an STFT time-frequency representation of the multichannel features, and outputs $T/5 \times N \times C \times 3$ vector coordinates, and $N$ is the assumed maximum co-occurring events, in our case $3$.

The input features are 4-channel 64-band log-mel spectrograms combined with SALSA-lite spatial features, all of which are truncated to include bins up to 9kHz, without mel-band aggregation following \cite{salsalite}).

Each model was trained on the training folds generated from the SRIR datasets described above.
In addition to this, data from the Sony-TAu Realistic Spatial Soundscapes 2022 (STARSS22) \cite{starss22} was added for training, using the 54 development sound mixtures for training, but witholding the remaining 52 clips for evaluation of the models, ensuring the results were exclusively on real recorded data from unseen rooms.
The models were trained for 100 epochs each,

\subsection{Evaluation}
\label{ssec:eval}
Evaluation was completed using join localization-detection metrics established in the DCASE 2020 challenge.
The detection metrics used were error-rate and F1 score for a spatial threshold within $20^\circ$ ($ER_{20^\circ}$ and $F_{20^\circ}$).
F1 score was macro-averaged to account for class distribution differences in the FSD50k audio subset used.
Localization metrics are class dependent localization error and recall ($LE_{\text{CD}}$ and $LR_{\text{CD}}$).

\section{Results}
\label{sec:results}
Despite the coarse physical approximations made by the ISM, the entirely synthetic SRIRs generated with this process performed nearly as well as the SRIRs recorded in real world settings.
Furthermore, the dataset of real SRIRs augmented by synthetic SRIRs outperformed both by a narrow margin, showing benefits of geometrical acoustics simulation for data augmentation for SELD tasks.

Regarding the cross-class average performance of the models in Table \ref{tab:avg_scores}, we can see that for our classification metrics, all three models perform relatively similarly, with a slight performance edge to the AUG-SRIR dataset trained models.
Looking into the per-class results in figure \ref{fig:perclassresults}, we can see that generally, all three models struggle with similar classes (telephone, laughter, door), but the AUG-SRIR dataset outperforms both in certain classes for which both other models perform poorly on (Water tap/Faucet, and Knock).

Looking at the localization based results, it appears that some amount of the performance differences between the SIM-SRIR trained models and the TAU-SRIR trained models can possibly be attributed to model fine-tuning; while SIM-SRIR trained models had poor localization performance on certain results (Water tap/Faucet, and Knock), the AUG-SRIR model outperformed the baseline TAU-SRIR dataset.
This suggests that the SIM-SRIR datasets are actually providing beneficial information for these sound classes missing from the TAU-SRIR datasets.
With more thorough model tuning, it's possible that the performance for SIM-SRIR trained models it even closer to that of the baseline.

\begin{table}[]
    \centering
    \begin{tabular}{c|cccc}
         & $ER_{20^\circ}$ & $F_{20^\circ}$ & $LE_{\text{CD}}$ & $LR_{\text{CD}}$\\
         \hline
        TAU-SRIR & \textbf{0.71} & 14.4\% & 55.1${^\circ}$ & 39.2\% \\
        SIM-SRIR & 0.73 & 13.0\% & 79.6${^\circ}$ & 34.8\%\\
        AUG-SRIR & 0.75 & \textbf{16.3}\% & \textbf{52.3${^\circ}$} & \textbf{42.3\%}
    \end{tabular}
    \caption{The cross-class evaluation metrics for models trained on data generated from different SRIR datasets}
    \label{tab:avg_scores}
\end{table}

\begin{figure}
    \centering
    \hspace*{-1cm}\includegraphics[width=0.45\textwidth]{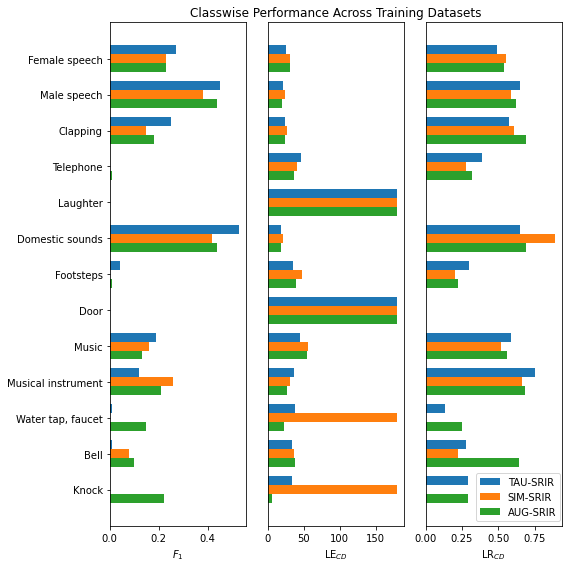}
    \caption{Per-class results of models trained on each dataset for F-measure within $20^{\circ}$, localization error, and localization recall.}
    \label{fig:perclassresults}
\end{figure}

\section{Conclusion}
\label{sec:conclusion}

In this work we demonstrated the potential of using acoustic simulation to generate spatial audio data for training SELD models.
We've shown that simulated SRIR data can improve the performance of SELD models as a form of data augmentation.
In addition to this, we've shown that simulated SRIRs, while not as effective as those recorded in real acoustic environments, can be used to effectively train SELD models, removing the relatively high cost of producing additional data for similarly performing results in a relatively limited setting.
Generating larger volumes of SRIRs over a wider range of acoustic conditions could provide even better results than these baselines, potentially demonstrating greater robustness over varying acoustic environments.
Furthermore, using a high-volume of simulated SRIRs to train a model, and using a hold-out of limited high-quality real-world data to fine the model could produce SoTA results.
This result is promising for future experiments involve SRIRs for use in acoustic simulation data.
Understanding the requirements for angular density in dynamic SRIR recordings can help inform future dataset collection practices, as well as the robustness of these models to noise; limited work was done exploring the effect of noise on the models trained with simulated SRIRs.
Further ablation studies are necessary to understand the limitations of geometrical acoustic methods for SELD-based tasks, but these early experiments suggest that these can provide a low-resource alternative to real-world SRIR recordings.

% -------------------------------------------------------------------------
% Either list references using the bibliography style file IEEEtran.bst
\bibliographystyle{IEEEtran}
\bibliography{refs}

\end{sloppy}
\end{document}